\documentclass[fleqn,usenatbib]{mnras}

\usepackage{newtxtext,newtxmath,multirow}

\usepackage{pdflscape}
\usepackage{booktabs}     
\usepackage{multirow}     
\usepackage{makecell}     
\usepackage{tabularx}     
\usepackage{siunitx}      
\usepackage{rotating}     
\usepackage{graphicx}     
\usepackage{caption}      
\usepackage{array}        
\usepackage{lineno}
\usepackage{ulem}

\usepackage[T1]{fontenc}
\usepackage{aecompl}
\newcommand{\eflux}{erg~cm$^{-2}$~s$^{-1}$}

\DeclareRobustCommand{\VAN}[3]{#2}
\let\VANthebibliography\thebibliography
\def\thebibliography{\DeclareRobustCommand{\VAN}[3]{##3}\VANthebibliography}

\usepackage{graphicx}	
\usepackage{amsmath}	
\usepackage{lscape}

\title[4FGL J1913.2+0512 nearby SS~433]{Temporal evolution of the periodic GeV signal from 4FGL J1913.2+0512 and analysis of the SS~433/W50 lobes}

\author[Ö. F. \c{C}oban et al.]{
Ömer Faruk \c{C}oban,$^{1}$\thanks{E-mail: coban@ice.csic.es}
D. F. Torres,$^{1,2,3}$
J. Li,$^{4,5}$
D. Hadasch,$^{1}$
A. De Sarkar$^{1}$ and 
M. Kerr$^{6}$
\\
$^{1}$ Institute of Space Sciences (ICE-CSIC), Campus UAB, Carrer 
de Can Magrans s/n, Barcelona 08193, Spain \\
$^{2}$ Institut d'Estudis Espacials de Catalunya (IEEC), Barcelona 08860, Spain \\ 
$^{3}$ Instituci\'o Catalana de Recerca i Estudis Avan\c{c}ats (ICREA), Barcelona 08010, Spain \\
$^{4}$ CAS Key Laboratory for Research in Galaxies and Cosmology, University of Science and Technology of China (USTC), Hefei 230026, China \\
$^{5}$ School of Astronomy and Space Science, University of Science and Technology of China (USTC), Hefei 230026, China  \\
$^{6}$ Space Science Division, Naval Research Laboratory, Washington DC 20375, USA
       }

\date{Accepted 2026 October 14. Received 2026 October 13; in original form 2026 April 27}

\pubyear{\the\year{}}

\begin{document}
\label{firstpage}
\pagerange{\pageref{firstpage}--\pageref{lastpage}}
\maketitle

\begin{abstract}
SS~433 is a microquasar whose relativistic jets precess every $\sim$162 days, providing a laboratory for jet-interstellar medium interactions.
We present a comprehensive analysis of 16 years of \textit{Fermi} Large Area Telescope data (August 2008--September 2024) of the SS~433/W50 field, using events in the $0.3-300$~GeV range and employing pulsar gating to mitigate contamination from the bright nearby pulsar PSR~J1907+0602.
We detect the GeV source 4FGL~J1913.2+0512 (TS = 45, where TS denotes the likelihood-ratio Test Statistic) with a power-law spectrum (photon index $2.61 \pm 0.08$) and confirm a GeV excess at the western lobe (TS = 17).
The eastern lobe of SS~433 is hinted at with lower significance.
One additional GeV excess, Fermi~J1909.6+0552 (TS = 20; TS = 28 over $0.1-300$~GeV), located outside the SS~433/W50 system, is revealed after gating.
Exposure-corrected Lomb-Scargle periodograms and precessional phase-folded light curves show a $\sim$162-day modulation in 4FGL~J1913.2+0512.
The phase-folded emission is mainly confined to the first half of the precessional cycle, while the source is not significantly detected in the second half. 
This periodic signal is mainly concentrated in the 5--10 yr interval, remains visible in the cumulative 0--10 yr data set, and is not recovered in the subsequent 10--16 yr interval.
Over the full 16-year data set, the modulation remains detectable but with reduced significance, consistent with dilution by the later non-modulated epoch.
These results indicate that the efficiency and/or geometry of gamma-ray production in the SS~433 environment evolves on multi-year timescales.
\end{abstract}

\begin{keywords}
gamma-rays: stars -- X-rays: binaries -- stars: individual: SS~433 -- ISM: jets and outflows
\end{keywords}



\section{Introduction}

SS~433 is a binary system with an orbital period of $13.087\pm0.003$ days, consisting of a compact object (likely a black hole) that accretes matter from a supergiant companion \citep{margon84,fabrika04,gies02}.
SS~433 is embedded in the supernova remnant W50, whose elongated shape has been sculpted by SS~433's bipolar jets \citep{elston87}. 
The system exhibits persistent jets with a velocity of $0.26c$ that precess around a cone with a period of $P_{\rm prec} = 162.250\pm0.003$ days \citep{davydov08}. The inner jets extend up to 0.1~pc from the central black hole \citep{blundell04}.
This precession causes the jets to trace a helical path through the surrounding medium. Doppler shifts of H and He lines in the optical band and highly ionized Fe lines in the X-ray band have been detected in the jets \citep{marshall02,migliari02}.

The jet termination lobes appear at a distance of $\sim$25 pc from the binary and extend out up to 100 pc \citep{brinkmann96,safiharb97,safiharb22,kayama22}.

Very-high-energy (VHE; $E > 100$~GeV) gamma rays have been detected from the jet lobes by the High Altitude Water Cherenkov (HAWC) observatory \citep{hawc18}. 
The High Energy Stereoscopic System (H.E.S.S.) has further revealed an energy-dependent morphology within these lobes. 
These findings are consistent with a leptonic scenario for the origin of the VHE gamma-ray emission \citep{hess24}.
More recently, the LHAASO Collaboration has reported an extended source emitting above 100~TeV, implying that particles are being accelerated up to PeV energies \citep{lhaaso25}. 

A number of early studies explored the GeV emission from SS~433 (e.g., see \citealt{bordas15, rasul19}), providing inconsistent results. \cite{fang20} and \cite{li20} analysed 10.5 years of \textit{Fermi} Large Area Telescope (LAT; \cite{atwood09}) data, gating off the nearby pulsar. The former work performed a joint analysis with the HAWC data focusing on the lobes, whereas the latter reported a hint of a GeV excess coincident with the western jet lobe (the “West Excess”) and focused on Fermi~J1913+0515 (now 4FGL~J1913.2+0512; \cite{ballet23}), a nearby source that had been included in the background model of \cite{fang20} but not considered as located at the binary or along the jets.
This source, which in projection lies at a distance of $\sim35$ pc away from SS~433, surprisingly presented a GeV periodicity with a period compatible with that of the jet precession. 

In this paper, we present a comprehensive analysis of 16 years of \textit{Fermi}-LAT gamma-ray observations of the SS~433 region, covering roughly 1.5 times the time span of the \cite{li20} study.
By leveraging a longer data set and also applying pulsar gating (to mitigate contamination from a nearby bright pulsar), we here perform a deeper search for GeV emission around SS~433 and test the appearance and eventual persistence of the periodic signal.
Our analysis is completely independent from that of \cite{li20}, and most of the relevant ingredients of the data analysis are different from those used back then (see below for details).
The primary aim of this work is therefore observational; constructing a detailed model for the origin of the periodic emission from 4FGL J1913.2+0512 is beyond the scope of the present work.

\begin{figure*}
\includegraphics[width=\textwidth]{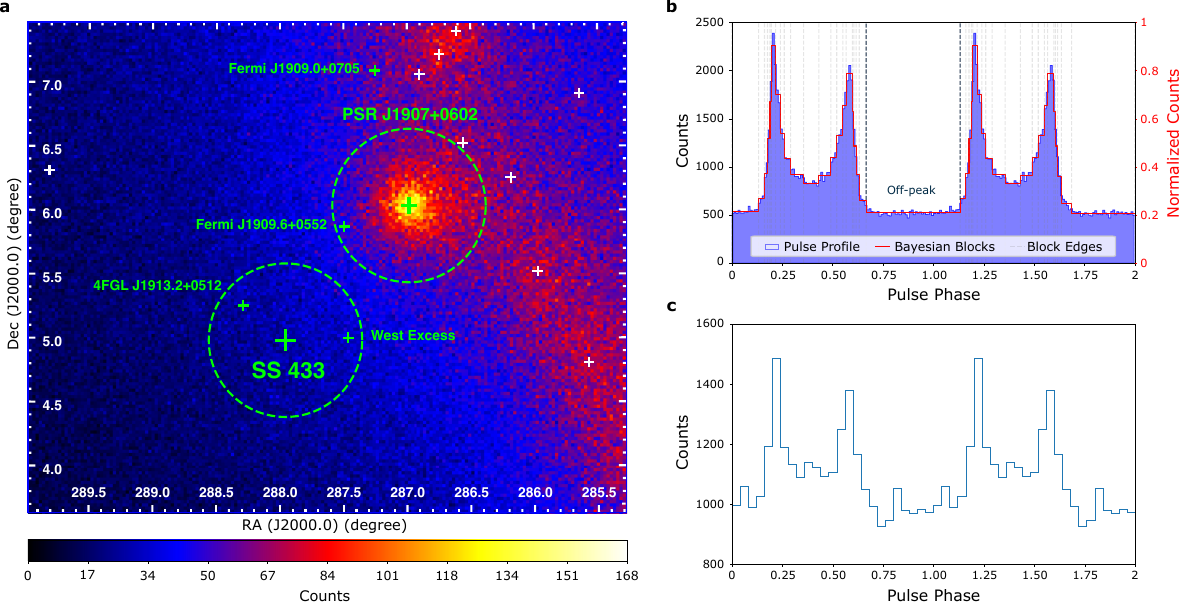}
\caption{\textbf{a)} Counts map of the Fermi-LAT field around the SS~433 region, constructed from events between $0.1-300$~GeV. White and green crosses indicate gamma-ray sources in the region, and dashed circles mark the regions used to produce pulse profiles. \textbf{b)} Pulse profile of events within 0.6\degr of PSR~J1907+0602 above 300~MeV. Two rotational pulse periods are shown with 100 phase bins per period. The off-peak interval is defined as $\phi = 0.666-1.132$ using Bayesian blocks. \textbf{c)} Pulse profile of photons above 300 MeV within 0.6\degr of SS~433, folded with the ephemeris of the nearby pulsar PSR~J1907+0602. Two rotational pulse periods are shown with 25 phase bins per period. This emphasizes the need to gate the pulsar off.}
\label{fig:PulsarContamination}
\end{figure*}

\section{Data and Analysis}

In this work, we use 16 years of \textit{Pass 8 Release 3 (P8R3) SOURCE} class data \citep{atwood13,bruel18} (from August 2008 to September 2024) collected by the LAT instrument onboard the \textit{Fermi} Gamma-ray Space Telescope, with reconstructed event energies ranging from 300~MeV to 300~GeV. 
The analysis was performed using the P8R3\_SOURCE\_V3 instrument response functions (IRFs). 
We selected a region of interest (ROI) with a radius of $15\degr$ centered on the optical position of the microquasar SS~433 at RA = 287.957$\degr$, Dec = 4.983$\degr$ \citep{wenger00}.
Photons with zenith angles larger than $90\degr$ were excluded to limit contamination from gamma rays produced in the Earth’s atmosphere. 
The selected events were binned spatially with a pixel size of $0.025\degr$ and divided into 30 logarithmically spaced energy bins.
The analysis was carried out using the binned maximum-likelihood method \citep{mattox96}.
The significance of a source detection was evaluated using the Test Statistic (TS), defined as twice the difference in the logarithm of the likelihoods between models with ($\mathcal{L}_1$) and without the source ($\mathcal{L}_0$): 
$   \mathrm{TS} = 2 \, \ln \left( {\mathcal{L}_1}/{\mathcal{L}_0} \right)$ \citep{mattox96}.
The detection significance can be approximated by $\sqrt{\mathrm{TS}}$ for one degree of freedom.
The spatial and spectral analysis was performed using the \texttt{FERMITOOLS} (v2.2.0) and \texttt{FERMIPY} packages (v1.3.1; \cite{wood17}). Energy dispersion corrections were applied to all sources except the isotropic diffuse component by adopting the \texttt{edisp\_bin=-2} setting.

The gamma-ray pulsar PSR~J1907+0602, with a spin period of $106.6$~ms \citep{abdo10}, is located only $1.4\degr$ from SS~433 and represents a significant source of contamination in this region.
Its bright gamma-ray emission dominates the surrounding field.
The LAT angular resolution is strongly energy-dependent, with a 68\% containment radius of roughly $2\degr$ at 300~MeV and about $1\degr$ at 1~GeV, improving progressively at higher energies.
To characterize its pulsed emission, we selected photons with energies above 300~MeV within a $0.6\degr$ radius centred on PSR~J1907+0602 (upper dashed circle in
Fig.~\ref{fig:PulsarContamination}a).
Rotational phases were assigned to the photons using an updated ephemeris for PSR~J1907+0602, following the methodology of \citet{kerr15}, implemented with
\textit{Tempo2} (v2023.05.1) and the \textit{Fermi} plug-in (v6.3) \citep{hobbs06,edwards06,ray11}. The resulting pulse profile is shown in Fig.~\ref{fig:PulsarContamination}b.
Using Bayesian blocks \citep{scargle13}, we identify the off-peak interval as $\phi=0.666-1.132$ in the two-cycle representation shown in the figure, corresponding to $\phi=0.666-1.0$ and $0.0-0.132$ when phases are restricted to the interval $0\leq\phi<1$.

To assess the contamination of the SS~433 region by PSR~J1907+0602, we repeated the phase analysis following \citet{li20}, using a $0.6\degr$ aperture centred instead on SS~433
(lower dashed circle in Fig.~\ref{fig:PulsarContamination}a). This aperture contains both 4FGL~J1913.2+0512 and the West Excess.
The photons were assigned rotational phases using the same PSR~J1907+0602 ephemeris. As shown in Fig.~\ref{fig:PulsarContamination}c, the pulsar modulation remains
clearly visible even in the aperture centred on SS~433, demonstrating that the field is substantially contaminated by PSR~J1907+0602.

We therefore restrict the subsequent analysis to the pulsar off-peak phases, $\phi=0.0-0.132$ and $0.666-1.0$. Accordingly, the prefactor parameters of all sources were scaled by $0.466$, corresponding to the fraction of the phase window used. 
Because the phase selection excludes the pulse peaks of PSR~J1907+0602, the phase-averaged catalog spectral model is not directly applicable to the gated data. We therefore retain a source component at the pulsar position and refit its residual off-peak emission with a power-law spectrum. The best-fit spectral parameters are reported in Table~\ref{tab:1}.

\begin{figure*}
\includegraphics[width=0.9\textwidth]{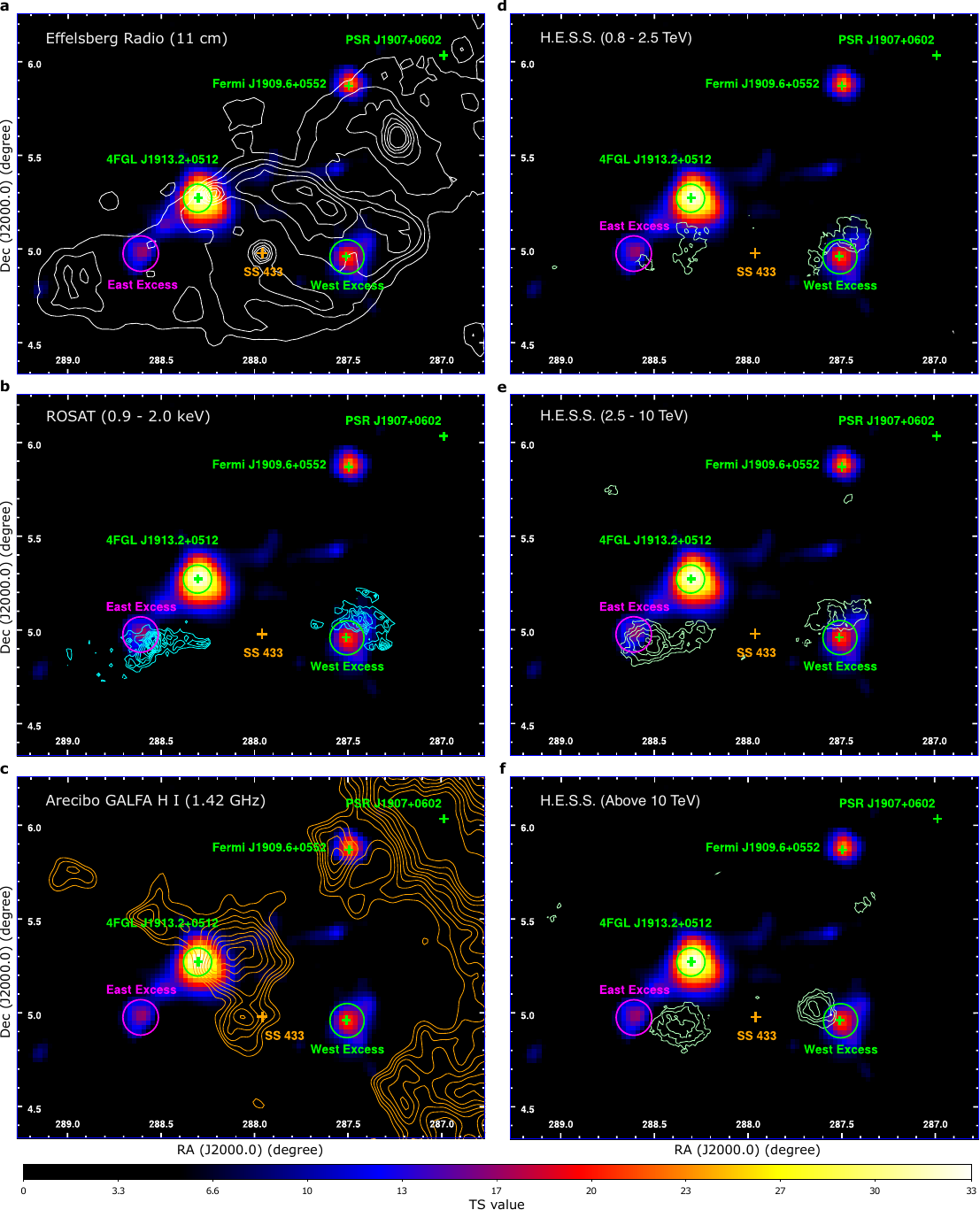}
\caption{\textit{Fermi}-LAT residual TS map of the SS~433 region during the off-peak phase of PSR~J1907+0602 in the $0.3-300$~GeV band. All 4FGL-DR4 sources are modelled and subtracted, except for the main source of interest, 4FGL~J1913.2+0512, which is excluded from the model to highlight its unmodelled emission. The $95\%$ confidence level localization regions of 4FGL~J1913.2+0512 and the Western Excess are indicated with green circles. \protect\textbf{a)} White contours are radio continuum from Effelsberg 11 cm survey from 300 mK with the increments of 200 mK. \protect\textbf{b)} Cyan contours show the X-rays observed with ROSAT in the 0.9-2 keV band. From $5\times10^{-5}$ counts $s^{-1}$ with increments of $1.67\times10^{-5}$ counts $s^{-1}$; \protect\cite{li20}. \protect\textbf{c)} Orange contours are Arecibo H I emission map integrated over $65-82$ km s$^{-1}$, scaled by $\sin|b|$ to enhance structures away from the Galactic plane \protect\citep{su18}. \protect\textbf{d--f)} H.E.S.S. TeV contours with increments of 1$\sigma$ starting from 3$\sigma$ in $0.8-2.5$~TeV (d), $2.5-10$~TeV (e), and above 10~TeV (f); \protect\citep{hess24}.}
\label{fig:TSMaps}
\end{figure*}

We constructed an initial source model including all known gamma-ray sources within $20\degr$ of SS~433 as listed in the 4FGL-DR4 catalog \citep{abdollahi22,ballet23}.
The diffuse background emission was modelled with the standard Galactic diffuse template \texttt{gll\_iem\_v07.fits}\footnote{\url{https://fermi.gsfc.nasa.gov/ssc/data/access/lat/BackgroundModels.html}} together with the isotropic diffuse component \texttt{iso\_P8R3\_SOURCE\_V3\_v1.txt} \citep{acero16}.
All spectral parameters of sources within $4\degr$ of SS~433 were left free during the fit, while more distant sources had their parameters fixed to catalog values. 
The source 4FGL~J1913.2+0512, which first appeared in DR2, is listed in DR4 as a Log-Parabola point source. 
Since above 300~MeV the Log-Parabola shape is not strongly preferred over a simple power-law, we modelled this source with a power-law spectrum.


We produced gamma-ray light curves for each source of interest using photon weighting (assigning each photon a weight based on its probability of originating from the source to maximize sensitivity; e.g., \cite{kerr19}).
We restricted the periodicity search to photon energies $E>1~\mathrm{GeV}$, where the point-spread function (PSF) is sharper and pulsar contamination is minimal. 
We constructed exposure-corrected, daily-binned light curves within $3\degr$ around 4FGL~J1913.2+0512, West~Excess, and PSR~J1907+0602.
These light curves were then analysed with the Lomb-Scargle (LS) algorithm \citep{lomb76,scargle82}, testing for any periodic component corresponding to a period of 162.25 days.

Using the precession period of 162.25 days and setting $T_0$ to the time of the largest separation of the moving emission lines in SS~433 (MJD 43507.9098; \cite{davydov08}), we folded all events with the \textit{gtpphase} timing package. 
The events were binned into half ($0.0-0.5$, $0.5-1.0$) and quarter ($0.00-0.25$, $0.25-0.50$, $0.50-0.75$, $0.75-1.0$) precessional phases. 
The ROI was fitted for each phase bin, and a precessional phase-resolved light curve was plotted from the resulting source fluxes.
Furthermore, to study the temporal evolution of the periodic signal, we divided the 16-year data set into: (i) the first 5 years (Aug~2008--2013), (ii) the next 5 years (Aug~2013--2018), (iii) the last 6 years (Aug~2018--Sep~2024), as well as two longer overlapping intervals, (iv) the first 10 years (Aug~2008--2018) and (v) the last 11 years (Aug~2013--Sep~2024) of the mission. For each subset, we generated LS periodograms and phase-resolved light curves for 4FGL~J1913.2+0512. This allowed us to assess whether the periodic modulation was continuously present or appeared and disappeared over time. 
To assess the likelihood of obtaining these signals by chance, we computed the false-alarm probability (FAP) levels using a bootstrap method. 
To do this, we constructed $10^{5}$ simulated light curves by assuming Gaussian white noise for the flux while keeping the temporal coordinates the same as in the actual light curve. 
We then computed $10^{5}$ LS periodograms from these simulated light curves and derived the FAP of detecting timing signals around our period of interest just as a noise fluctuation.

All results were checked for statistical robustness. To estimate spectral systematic uncertainties, we varied the normalization of the Galactic diffuse emission model by $\pm6\%$ and adopted the bracketing effective area (Aeff) method\footnote{\url{https://fermi.gsfc.nasa.gov/ssc/data/analysis/scitools/Aeff_Systematics.html}} to assess the impact of instrument response uncertainties. The contribution of the latter to the uncertainty was found to be negligible compared to that from the Galactic diffuse component. In addition, to evaluate any influence on the 4FGL~J1913.2+0512 results, we performed several robustness tests, including changing the ROI size to $10\degr$ and including/excluding nearby sources. We found that none of these tests or systematic variations significantly affect our conclusions.

\section{Results}
\label{sec:Results}

\subsection{Gamma-ray Sources in the SS~433 Region}
\label{ssec:Sources}

Our analysis confirms the presence of significant GeV gamma-ray emission in the vicinity of SS~433.
Fig~\ref{fig:TSMaps} illustrates the $0.3-300$~GeV residual TS map of the SS~433 region using 16 years of off-peak data.
Three prominent gamma-ray excesses are evident: the previously known source 4FGL~J1913.2+0512 to the northeast of SS~433, the “West Excess” cospatial with the western lobe, and Fermi~J1909.6+0552 to the northwest of the system.
The best-fit position of 4FGL~J1913.2+0512 is localized at RA = $288.31\degr \pm 0.03\degr$, Dec = $5.27\degr \pm 0.03\degr$, in agreement with coordinates reported by \cite{li20}. 
Assuming a power-law spectral shape, the source is significantly detected with a TS value of 45, a photon index of $\Gamma=2.61\pm0.08_{\mathrm{stat}}\pm0.03_{\mathrm{sys}}$, corresponding to a flux of $(7.25\pm0.81_{\mathrm{stat}}\pm1.92_{\mathrm{sys}})\times10^{-12}$~\eflux above 300~MeV. 
The source is spatially coincident with a clump of interstellar gas (seen in H I contours; see Fig.~\ref{fig:TSMaps}c) adjacent to the northeastern edge of W50 \citep{li20,lhaaso25}. Furthermore, the distance to the atomic cloud at this location is consistent with that of SS~433.

The West Excess, coincident with the western jet termination lobe and with X-ray/radio emission, is localized at RA = $287.50\degr \pm 0.04\degr$, Dec = $4.96\degr \pm 0.04\degr$ in agreement with the coordinates reported by \cite{li20}. The likelihood analysis of the west GeV excess yields a TS value of $17$ (below the formal source-detection threshold of 25), a spectral index of $\Gamma=2.46\pm0.18_{\mathrm{stat}}\pm0.25_{\mathrm{sys}}$, and an energy flux of $(4.59\pm0.85_{\mathrm{stat}}\pm2.55_{\mathrm{sys}})\times10^{-12}$~\eflux above 300~MeV.

Additionally, the 16-year data set reveals a hint of GeV emission from the eastern jet termination lobe, hereafter referred to as the “East Excess”.
When an additional source is included at that position, the fit yields TS = 7, driven by a few photons in the $20-40$~GeV range. 
We derive a 95\% confidence upper limit on the energy flux of the East Excess of $1.56\times 10^{-12}$ \eflux.
However, due to its low significance and close proximity to 4FGL~J1913.2+0512 (within < 0.5\degr), it is not possible to reliably model its emission. 
We tested whether including this source in the model affected any of the periodicity results, and found that it does not produce any change.
Therefore, we continued the analysis without including this source in the model, while remarking on its spatial coincidence with the eastern lobe of SS~433.

In addition, the residual TS map reveals a new gamma-ray source not present in the 4FGL-DR4 catalog. Fermi~J1909.6+0552, lies roughly $1\degr$ northwest of SS~433, slightly north of the jet axis. The source is spatially coincident with an atomic cloud concentration of similar size (see Fig.~\ref{fig:TSMaps}c). When modelled as a power-law source, it yields a TS value of 20, at flux $(5.16\pm1.07_{\mathrm{stat}}\pm2.15_{\mathrm{sys}})\times10^{-12}$~\eflux, with spectral index of $\Gamma=2.26\pm0.09_{\mathrm{stat}}\pm0.20_{\mathrm{sys}}$. The localized position is RA = $287.50\degr \pm 0.03\degr$, Dec = $5.87\degr \pm 0.03\degr$. To test whether the detection is significant, we repeated the maximum likelihood analysis between $0.1-300$~GeV, obtaining a TS of 28, above the conventional threshold of 25.
Table~\ref{tab:1} summarizes the spectral fitting results for these sources with statistical and systematic uncertainties.

\begin{figure*}
\includegraphics[width=\textwidth]{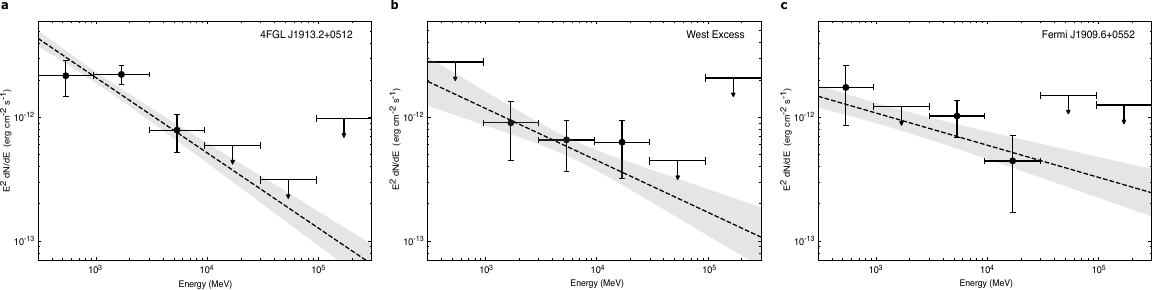}
\caption{SEDs of \textbf{(a)} 4FGL~J1913.2+0512, \textbf{(b)} West Excess, and \textbf{(c)} Fermi~J1909.6+0552 in the $0.3-300$~GeV energy range. The black points represent the \textit{Fermi}-LAT measurements with 1$\sigma$ statistical uncertainties. Each data point is obtained from an independent fit performed in the corresponding energy bin, with the photon index fixed to the value obtained from the global fit and only the source normalization left free. Downward arrows indicate $95\%$ confidence-level upper limits for energy bins with $\mathrm{TS} < 4$. The dashed lines show the best-fitting power-law models obtained from fitting the full $0.3-300$~GeV energy range, while the gray shaded bands represent the corresponding statistical uncertainties derived from the covariance matrices of the fitted spectral parameters.}
\label{fig:Sed}
\end{figure*}

We extract spectral energy distributions (SEDs; $0.3-300$~GeV) for 4FGL~J1913.2+0512, West Excess, and Fermi~J1909.6+0552. Representative SED panels are shown in Fig. \ref{fig:Sed}. The data points show the fluxes obtained from independent likelihood fits in each displayed energy interval, whose boundaries are indicated by the horizontal bars. In these bin-by-bin fits, the photon index was fixed to the value obtained from the global fit, and only the source normalization was left free. The dashed curve shows the best-fitting spectral model obtained from the global likelihood fit over the full energy range, and the gray shaded band represents the statistical uncertainty derived from the covariance matrix of the fitted spectral parameters. The spectra are well described by power laws over the LAT band; we do not find compelling evidence for curvature, although limited statistics in some bands preclude strong constraints.

\begin{center}
 \begin{table*}
  \caption{Best-fit spectral and spatial model parameters for the analysed \textit{Fermi}-LAT sources above 300 MeV. }
  \label{tab:1}
  \begin{tabular}{cccccccc}
    \hline
    \hline
    Source Name & Spatial Model & Spectral Model & Flux                     & Index  & TS & RA & Dec \\
                &               &                & ($10^{-12}$~erg~cm$^{-2}$~s$^{-1}$) &$(\Gamma)$       &    & $(\degr)$   & $(\degr)$    \\
    \hline
    4FGL~J1913.2+0512 & Point Source & Power-Law & $7.25\pm0.81_{\mathrm{stat}}\pm1.92_{\mathrm{sys}}$ & $2.61\pm0.08_{\mathrm{stat}}\pm0.03_{\mathrm{sys}}$ & 45 & $288.31\pm0.03$ & $+5.27\pm0.03$\\
    West Excess & Point Source & Power-Law & $4.59\pm0.85_{\mathrm{stat}}\pm2.55_{\mathrm{sys}}$ & $2.46\pm0.18_{\mathrm{stat}}\pm0.25_{\mathrm{sys}}$ & 17 & $287.50\pm0.04$ & $+4.96\pm0.04$ \\
    Fermi~J1909.6+0552 & Point Source & Power-Law & $5.16\pm1.07_{\mathrm{stat}}\pm2.15_{\mathrm{sys}}$ & $2.26\pm0.09_{\mathrm{stat}}\pm0.20_{\mathrm{sys}}$ & 20 & $287.50\pm0.03$ & $+5.87\pm0.03$\\
    4FGL~J1907.9+0602 & Point Source & Power-Law & $15.11\pm1.31_{\mathrm{stat}}\pm4.33_{\mathrm{sys}}$ & $2.88\pm0.07_{\mathrm{stat}}\pm0.07_{\mathrm{sys}}$ & 46 & $286.98\pm0.02$ & $+6.04\pm0.02$ \\
    \hline
  \end{tabular}
 \end{table*}
\end{center}

\subsection{Periodic modulation}

\begin{center}
\begin{table*}
\caption{Spectral fit parameters for 4FGL~J1913.2+0512 across different time spans.}
  \label{tab:2}
\begin{tabular}{clccccc}
\hline
\hline
\multicolumn{1}{l}{}     &                                                      & \multicolumn{1}{c}{$0-5$ yrs}  & \multicolumn{1}{c}{$5-10$ yrs} & \multicolumn{1}{c}{$10-16$ yrs} & \multicolumn{1}{c}{$0-10$ yrs} & \multicolumn{1}{c}{$0-16$ yrs} \\ \hline
\multirow{3}{*}{300~MeV to 300~GeV} & TS                                                   &7                           &30                            &11                 &34                          &45                             \\
                         & Photon Index                                         &...               &$2.56\pm0.08$                 &$2.64\pm0.76$           &$2.58\pm0.17$               &$2.61\pm0.08$ \\
                         & Flux ($10^{-12}$ \eflux)                             &$10.28~(95\%~\mathrm{UL})$               &$9.92\pm0.98$                 &$6.27\pm1.81$          &$7.56\pm1.58$               &$7.25\pm0.81$                             \\ \hline
\multirow{4}{*}{1 to 300~GeV}   & TS                                                   &4                           &30                            &19                 &29                          &45           \\
                         & Photon Index                                         &...                         &$3.04\pm0.02$                 &$3.65\pm0.47$           &$2.77\pm0.02$               &$3.01\pm0.21$       \\
                         & Flux ($10^{-12}$ \eflux)                             &$3.97~(95\%~\mathrm{UL})$                         &$5.24\pm0.21$                 &$3.79\pm1.01$           &$3.86\pm0.18$               &$3.75\pm0.66$       \\
\hline
\end{tabular}
\end{table*}
\end{center}

\begin{figure*}
\includegraphics[width=\textwidth]{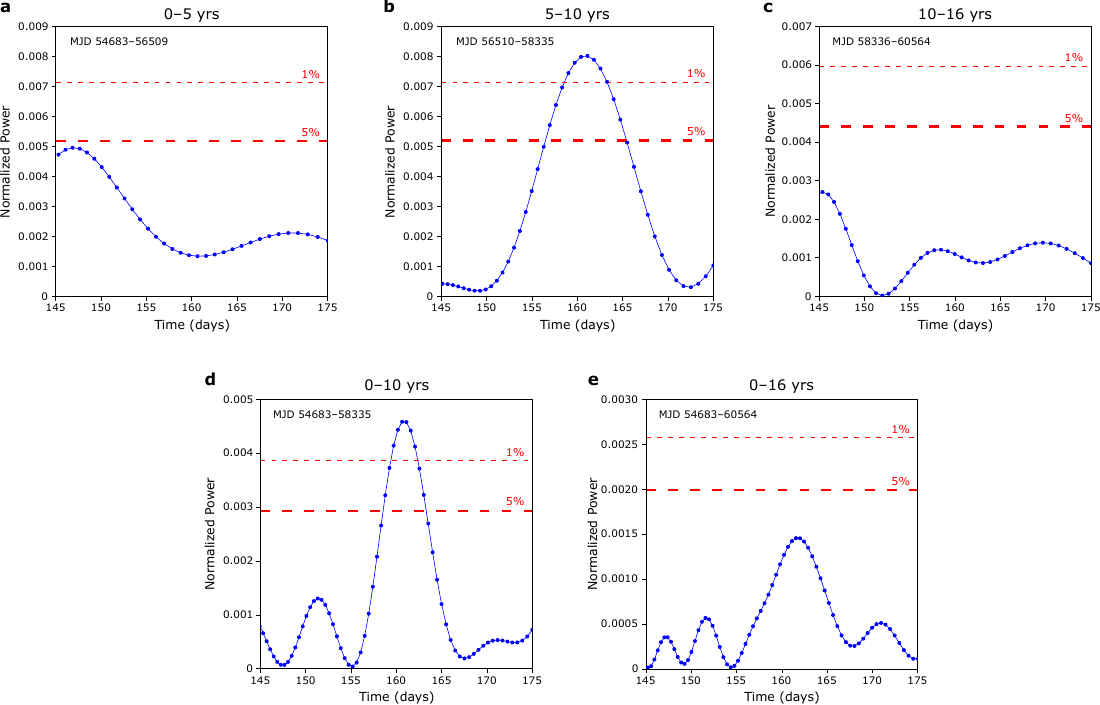}
\caption{Exposure-corrected Lomb-Scargle periodograms constructed from the $1-300$~GeV weighted light curve of 4FGL~J1913.2+0512 over different time intervals: \textbf{(a)} 5 years, \textbf{(b)} $5-10$ years, \textbf{(c)} $10-16$ years, \textbf{(d)} 10 years, and \textbf{(e)} 16 years. The blue curve represents the normalized power, while the red dashed lines indicate the $1\%$ and $5\%$ false-alarm levels computed over the 145–175 day range, accounting for the number of independent trials within this interval.} 
\label{fig:LSPeriodograms}
\end{figure*}

4FGL~J1913.2+0512 stands out as the only source showing any evidence of a periodic signal. Fig.~\ref{fig:LSPeriodograms} shows the evolution of the LS periodogram in five-year intervals, as well as for cumulative combinations of these windows, up to the total extent of our observations. 
A notable outcome of the extended data set is that the gamma-ray emission from 4FGL~J1913.2+0512 varies on multi-year timescales (see Table~\ref{tab:2} and Fig.~\ref{fig:LSPeriodograms}).
The source shows no significant emission during the early years of the \textit{Fermi} mission, but later becomes brighter and exhibits a clearly detectable periodic modulation during the subsequent five years.

During the first decade (2008--2018), the periodic signal is clearly present: the LS periodogram displays a $\sim$162-day peak with a FAP $< 1\%$, and the phase-folded light curve shows a pronounced on/off modulation between the two half-cycles.
This is consistent with the findings of \cite{li20}. 
In the later years (2018--2024), 4FGL~J1913.2+0512 is detected with lower significance and the $\sim$162-day modulation associated with jet precession is no longer visible in the LS periodogram (Fig.~\ref{fig:LSPeriodograms}). A $\sim$30\% flux reduction alone is insufficient to remove the periodic peak (See appendix~\ref{app:SimulatedLS}).

\begin{figure*}
\includegraphics[width=\textwidth]{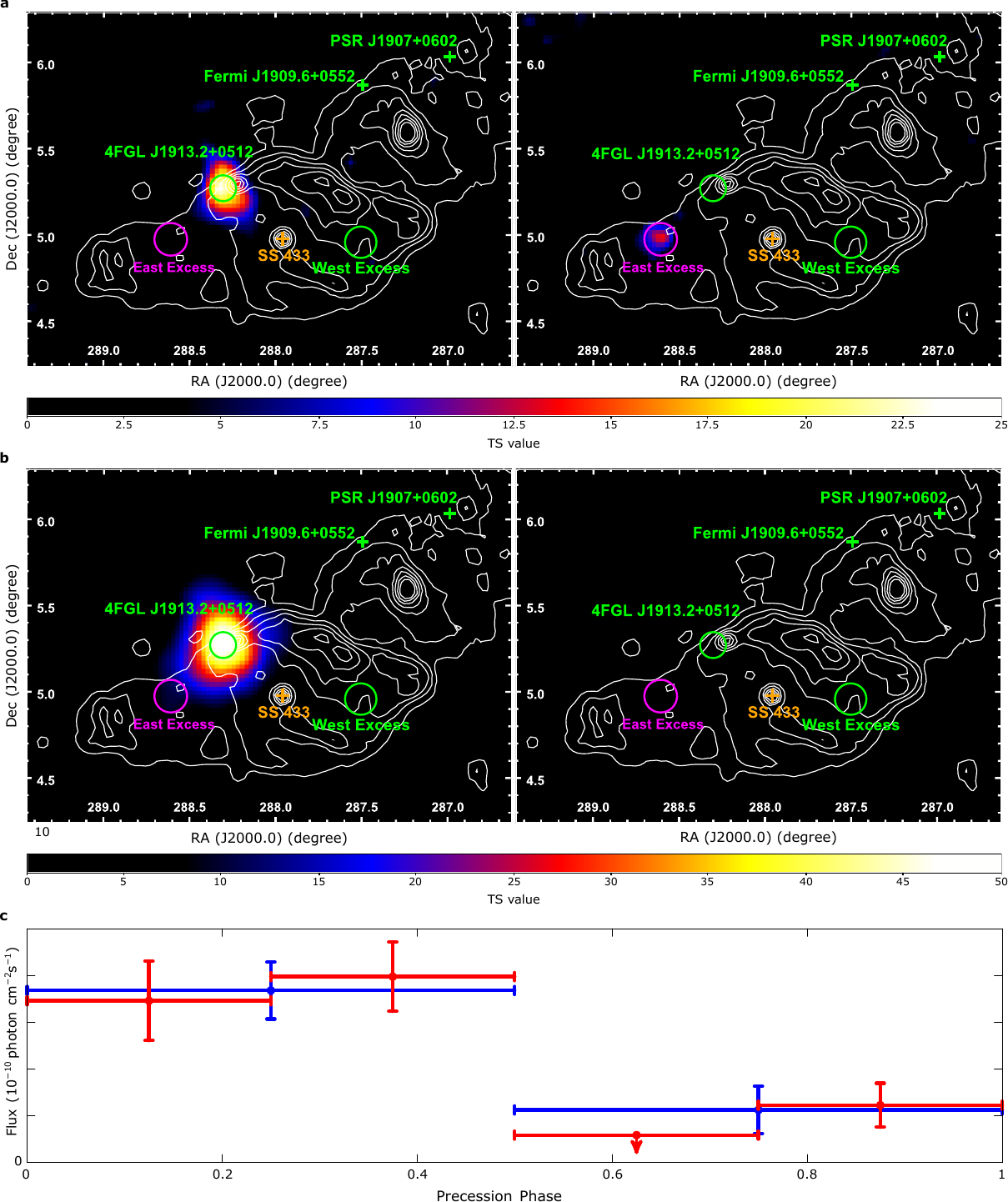}
\caption{\textbf{a)} Phase-folded TS Maps of the SS~433 region in the $1-300$~GeV band in the precessional phases $0.0-0.5$ (left) and $0.5-1.0$ (right), computed with 4FGL~J1913.2+0512 excluded in the model and a test-source photon index $\Gamma=2.0$. The East Excess is produced by only three photons; see the text for further discussion. \textbf{b)} Same as panel \textbf{a}, but using $\Gamma_{\rm ts}=3.2$, corresponding to the best-fit photon index of 4FGL~J1913.2+0512 in the bright interval ($0.0-0.5$). \textbf{c)} Precessional phase resolved light curve of 4FGL~J1913.2+0512 in the $1-300$~GeV band. Blue and red data points correspond to bin sizes of 0.5 and 0.25, respectively. Vertical error bars indicate 1$\sigma$ statistical uncertainties, and downward arrows represent $95\%$ confidence level upper limits (TS$~< 4$).}
\label{fig:PhaseTSandLC}
\end{figure*}

To examine the precessional dependence of the emission, we produced the precessional phase-folded TS maps and
light curve of 4FGL~J1913.2+0512 (Fig.~\ref{fig:PhaseTSandLC}), combining all 16 years of data. The TS maps are shown for two choices of the test-source photon index: $\Gamma_{\rm ts}=2.0$ (Fig.~\ref{fig:PhaseTSandLC}a) and $\Gamma_{\rm ts}=3.2$ (Fig.~\ref{fig:PhaseTSandLC}b), where $\Gamma_{\rm ts}=3.2$ corresponds to the best-fit index of 4FGL~J1913.2+0512 in the brighter ($\Phi_{\rm prec}=0.0$-0.5) interval.

The two TS maps are broadly consistent, but an apparent "East Excess" is visible in the $\Gamma_{\rm ts}=2.0$ map (Fig.~\ref{fig:PhaseTSandLC}a). This feature is driven by a very small number of hard events: modelling the excess yields a hard spectrum ($\Gamma\sim1.7$), and it is produced by only three photons with $E>20$~GeV, all of which fall in the second half of the precessional phase. A harder assumed test-source index up-weights high-energy photons, making these few events appear comparatively prominent in the $\Gamma_{\rm ts}=2.0$ TS map; however, the same excess is not visible in the TS map computed with $\Gamma_{\rm ts}=3.2$ (Fig.~\ref{fig:PhaseTSandLC}b). This should not be considered evidence for periodicity at this stage, but rather a consequence of the random distribution of only three events within half of the precessional cycle.

The phase-folded TS maps and light curve (Fig.~\ref{fig:PhaseTSandLC}) show that the gamma-ray emission is not uniform across the $0 \leq \Phi_{\rm prec} < 1$ phase: it is brighter during $\Phi_{\rm prec} \approx 0.0$-0.5 and much fainter during $\Phi_{\rm prec} \approx 0.5$-1.0. When dividing the data into these two intervals, the source appears to be essentially quiet during the latter half of the cycle, in agreement with the findings of \cite{li20}.
The phase-folded flux (two bins per cycle in Fig.~\ref{fig:PhaseTSandLC}c) shows nearly zero emission in the fainter half-cycle, while the brighter half-cycle accounts for nearly all of the observed flux.
We further divided the phase light curve into four bins (red points; Fig.~\ref{fig:PhaseTSandLC}c) to search for finer structure. However, given the limited statistics per bin, it is difficult to construct a detailed pulse profile. A simple step function between an “on” state (first half) and an “off” state (second half) provides an adequate description of the data. Neither the West Excess nor the other nearby sources exhibit comparable modulation when their data are folded on 162 days; their phase-binned fluxes are constant within errors.

\begin{figure}
\includegraphics[width=\columnwidth]{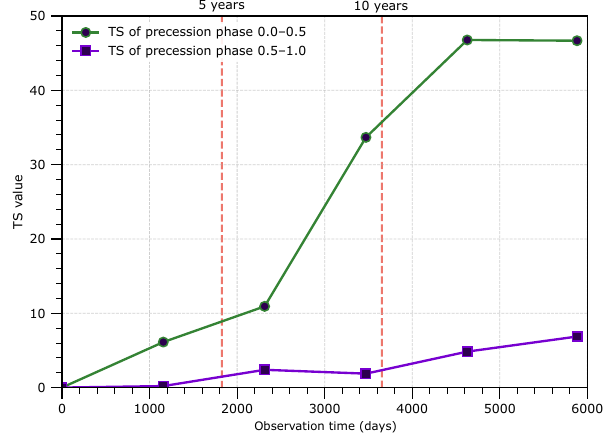}
\caption{ TS evolution of 4FGL~J1913.2+0512 in the $1-300$~GeV band for precessional phases $0.0-0.5$ (green) and $0.5-1.0$ (purple).}
\label{fig:CumulativeTS}
\end{figure}

The temporal evolution of the LS periodogram (Fig.~\ref{fig:LSPeriodograms}) is compatible with that of the cumulative source TS for the bright ($0.0-0.5$) and faint ($0.5-1.0$) precession half-cycles (Fig.~\ref{fig:CumulativeTS}).
During the first five years, the source is not significantly detected in either half-cycle; the TS difference between the bright and faint phases is small, and no periodic signal is observed. In contrast, during the second 5-year interval, a strong periodic signal emerges in the LS periodograms. This coincides with a sharp increase in TS in the bright phase ($0.0-0.5$) (Fig.~\ref{fig:CumulativeTS}), while the faint phase remains weak. These trends suggest that the rise in TS is driven predominantly by photons arriving within a specific precessional phase interval, naturally leading to the detection of precessional modulation during this epoch.
In the final six years, the TS difference between the two half-cycles stops growing and remains approximately constant; accordingly, the LS periodograms do not show a periodic signal. Considering the full 16-year data set, a TS difference between the two half-cycles persists, but it is not substantially larger than that observed over the first 10 years. Consequently, although a periodic signal remains detectable, its significance is reduced by dilution from the later, non-modulated epoch and by additional background photons.
This behaviour indicates that the appearance and subsequent disappearance of the periodicity are driven primarily by the evolution of the phase-dependent emission, rather than by a long-term change in the source flux.

\section{Discussion}

We note that the mission has no known temporal effects corresponding to a 162-day period arising from the spacecraft's orbit or its sky-scanning strategy\footnote{\url{https://fermi.gsfc.nasa.gov/ssc/data/analysis/LAT_caveats_temporal.html}}. The closest known periodicities are a 91-day modulation caused by the LAT's PSF symmetry interacting with the satellite's annual roll, and a 1-year variation linked to the Sun's apparent motion for sources near the ecliptic. Neither of these temporal artifacts is close to 162 days, and no known systematic effect is expected to produce variability at this timescale.
In any case, should the 4FGL J1913.2+0512 periodic signal be the result of a systematic effect at the mission level, there would a-priori be no reason why other sources nearby in the same ROI would not show it as well. This has been confirmed to be the case also in the period where the signal concentrates, from 5 to 10 years after launch.

To investigate the further possibility of artificial signals, we produced (i) LS periodograms of the \textit{Fermi}-LAT exposure function and (ii) a window-function analysis using the arrival times of all photons within the full ROI. The window-function analysis characterizes the temporal sampling of the LAT events, which is influenced by the spacecraft’s scanning strategy, orbital motion, and data gaps. It quantifies how these observational patterns can modulate photon arrival times and imprint artificial periodicities in the data.
Neither of these diagnostics showed any power near 162 days for the time intervals during which the periodicity is significantly detected, indicating that the modulation in 4FGL~J1913.2+0512 does not arise from exposure or spacecraft systematics. See appendix~\ref{app:Exposure} for a daily exposure history analysis.

In comparison with \cite{li20}, this analysis uses a newer source catalog (4FGL-DR4, based on 14 years of \textit{Fermi}-LAT data, which contains about $42\%$ more sources than the initial 4FGL catalog based on 8 years), a newer isotropic diffuse component (\texttt{iso\_P8R3\_SOURCE\_V3\_v1.txt} instead of \texttt{iso\_P8R3\_SOURCE\_V2\_v1.txt}), an updated IRF (P8R3\_SOURCE\_V3 instead of P8R3\_SOURCE\_V2), model a new source that was not taken into account before (Fermi~J1909.6+0552), uses an improved spatial binning ($0.025\degr$ vs $0.1\degr$), an updated and extended pulsar ephemeris to gate off the emission from PSR J1907+0602, and recomputes the Bayesian blocks leading to a slightly different off-peak interval ($\phi = 0.666-1.132$ vs $\phi = 0.697-1.136$).
The appearance of the same periodic signal (same period, at the same time interval of the analysis, with comparable significance) under a different 
analysis setup provides further confidence in its non-spurious character.

Fig.~\ref{fig:LSPeriodograms} shows that the timing signal is mostly found in the 5 to 10 years interval since launch, with little power outside this range.
To verify this, we performed a sliding-window analysis with a window size of 5 years and a step of 1-year since the start of the mission.
Taking into account that the mean level, the noise variance, or the presence of additional slowly varying components differs across successive data segments, it may alter the period signaled as well as the LS periodogram normalization. 
The FAP was recomputed for each period individually to account for these effects in a consistent manner.
The window with the most power across the mission life-time is indeed the second five years.
Moving the sliding window years before or after that window decreases the signal.

The present work  provides an observational re-examination of the reported modulation and studies its temporal evolution. We do not aim here to produce a detailed model of the physical process that gives rise to the phenomenology found.
Particles from SS~433 appear to have travelled for tens of parsecs without losing coherence, and produced a signal along this period.
\cite{li20} considered hadronic emission from protons supplied by SS 433 and interacting with the gas enhancement towards 4FGL J1913.2+0512. 
Protons accelerated in the jets or supplied by the equatorial outflow could account for the mean gamma-ray flux, but isotropic diffusion over tens of parsecs would smear out the precessional modulation.
Direct illumination by the precessing jet was also considered unlikely because the source lies outside the extrapolated jet cone and is separated from the eastern jet termination region by tens of parsecs in projection as well. 
The equatorial outflow may offer a more favourable geometry, although it does not by itself resolve the propagation problem. 
These authors also discussed anisotropic diffusion or streaming along a magnetic flux tube.
This requires efficient delivery of protons while maintaining a sufficiently narrow spread of arrival times and a short residence time in the target. 
Any substantial broadening of the arrival times would suppress the modulation. 
A larger diffusion coefficient would alleviate the temporal-coherence requirements; a complete scenario still remains to be presented.

The temporal modulation shows that the signal is not stable.
Whatever the mechanism generating the emission would ultimately be, it would seem that both geometrical and physical factors can easily conspire against maintaining the signal stable over time.
We therefore interpret the loss of periodicity as a possible outcome of an evolution in the efficiency and/or geometry of the particle transport that feeds this gamma-ray production and affects the coherence of the precessional modulation. 
Nevertheless, no strong conclusion can be drawn with the current understanding of the system without a specific modelling approach.

We caution that the apparent ``East Excess'' in Fig.~\ref{fig:PhaseTSandLC}a is driven by only three $E>20$~GeV photons falling in the $\Phi_{\rm prec}=0.5$--1.0 half-cycle. Therefore, we do not consider it evidence for periodicity, but a chance clustering of a few events within half of the precessional cycle.

Fermi~J1909.6+0552 is spatially coincident with an atomic gas concentration (Fig. \ref{fig:TSMaps}c). This coincidence suggests cosmic-ray interactions with the gas as a possible origin of the gamma-ray emission, although its physical origin remains uncertain.

\begin{figure*}
\includegraphics[width=0.99\textwidth]{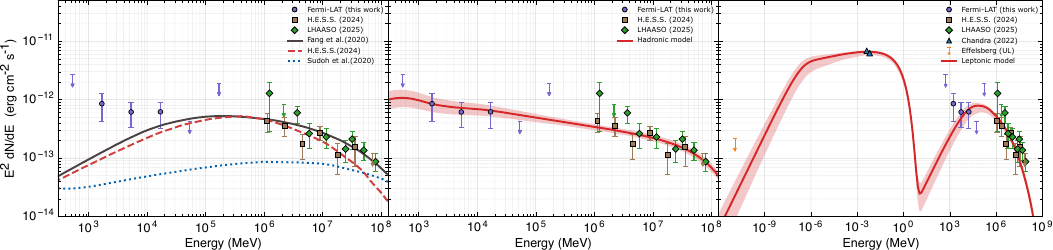}
\caption{Multiwavelength SED of the West Excess obtained with \protect\textit{Fermi}-LAT (this work), H.E.S.S. \protect\citep{hess24}, and LHAASO \protect\citep{lhaaso25}. Vertical bars represent 1$\sigma$ uncertainties, while downward arrows indicate 95$\%$ confidence level upper limits. The left panel shows, for reference, model curves from \protect\cite{fang20}, \protect\cite{hess24}, and \protect\cite{sudoh20}.  The middle and right panels show the best-fit model gamma-ray SEDs obtained from hadronic and leptonic scenarios, respectively, along with data points considered from the analysis of \protect\textit{Fermi}-LAT (this work), H.E.S.S. \protect\citep{hess24}, and LHAASO \protect\citep{lhaaso25}. The right panel additionally includes radio and X-ray measurements from Effelsberg \protect\citep{reich84, reich90, furst90} and \protect\textit{Chandra} \protect\citep{kayama22}, respectively. Both of these model SEDs also contain 1$\sigma$ error bands in red shaded region. Details are given in the text.}
\label{fig:MW_SED}
\end{figure*}

Fig.~\ref{fig:MW_SED} shows the SED of the western lobe of SS~433 as obtained in this work, together with multi-TeV, X-ray and radio measurements. 
The H.E.S.S. TeV data points were considered from \cite{hess24}.
The radio flux from the 11 cm Effelsberg survey \citep{reich84, reich90, furst90} was estimated from the same spatial region from which the TeV data was extracted.
Similarly, the Chandra X-ray fluxes were also extracted from the same TeV-emitting region \citep{kayama22}, while taking into account the small field-of-view of Chandra by considering a correction factor of 10$\%$ as derived from surface brightness of the jets in soft X-ray \citep{brinkmann07}.
The LHAASO data points for the western lobe were considered from that explicitly reported by LHAASO collaboration \citep{lhaaso25}.
The {\it Fermi}-LAT GeV data points were used from our analysis of the western excess.
Here, note that although the centroid of the GeV western excess reported in this paper is not exactly coincident with the H.E.S.S. emission centroid, the TeV significance contours partially overlap with the 95$\%$ confidence localization region of the western excess, which is similar to that previously found in \cite{fang20} and \cite{hess24}.
While the GeV spectrum used by previous modelling efforts was largely represented by upper limits, in this paper, we provide and subsequently use statistically significant GeV flux data for the western excess in our multiwavelength SED discussed below.

Several theoretical models have been proposed to explain the TeV emission recently, including those by \cite{fang20, sudoh20}, and \cite{hess24}. 
In these models, electrons accelerated at the jet termination shock, or at the base of the jets, produce synchrotron (radio--X-ray) and inverse-Compton (IC; GeV--TeV) emission on CMB and Galactic IR photon fields.
However, all of these models underpredict the LAT flux in the GeV range.

Here, we aim to explain the observed multiwavelength data simultaneously, for which we have used the \texttt{GAMERA} code \citep{hahn15, hahn22}.
We start by discussing a hadronic scenario. 
Due to an absence of dense molecular clouds at the position of the western lobe, we assume an ambient number density of $n \approx 1 \ \rm cm^{-3}$, similar to that of the interstellar medium.
We consider the source distance of $\sim$ 5.5 kpc \citep{blundell04}, and an age of $\sim 2.2 \times 10^4$ yrs \citep{goodall11, li20}. 
The injection spectrum of the accelerated protons is considered to follow a simple power-law of the form $Q_{\rm p}(E_{\rm p}) = Q_{\rm p, 0}(E_{\rm p}/1 \ \rm TeV)^{-{\alpha_p}}$ in the energy range of $E_{\rm p, \ min} \approx 1 \ \rm  GeV$ ($\sim$ proton rest mass energy) and $E_{\rm p, \ max} \approx 1 \ \rm PeV$.  
The total proton luminosity can be found through the relation $L_{\rm p} = \int^{E_{\rm p, max}}_{E_{\rm p, min}}E_{\rm p} \ Q_{\rm p}(E_{\rm p}) \ dE_{\rm p}$.
We then allow the proton spectral index $\alpha_p$ and total proton luminosity $L_{\rm p}$ to vary freely so that the model spectrum fits the combined GeV--TeV SED, which leads to the best-fit values of $\alpha_{\rm p} = 2.21 \pm 0.04$ and $\mathrm{log_{10}}(L_{\rm p}/ \rm erg \ s^{-1}) = 38.39 \pm 0.12$.
Fig. \ref{fig:MW_SED} shows the best-fit hadronic model SED, along with a 1$\sigma$ error band, plotted against the observed spectrum.
It can be seen that this model yields a $\pi^0$-decay gamma-ray spectrum that reproduces the observed SED well.
Furthermore, the best-fit value of the required proton luminosity $L_{\rm p}~(\approx 2.5 \times 10^{38} \ \rm erg \ s^{-1}$) is consistent with the typically assumed $10\%$ cosmic ray (CR) conversion efficiency from the jet kinetic luminosity of $L_{\rm jet} \approx 3 \times 10^{39} \ \rm erg \ s^{-1}$ \citep{li20}.
The best-fit proton spectral index is also consistent with that typically estimated from diffusive shock acceleration occurring at shocks.
Taken together, these results indicate that a hadronic origin for the observed gamma-ray emission from the western lobe is plausible.
However, it is to be noted that a pure hadronic scenario will be unable to explain the X-ray/radio emission from the western lobe of SS~433.

Alternatively, we have also explored a leptonic interpretation for the observed multiwavelength emission, where the gamma-ray data is explained via the IC upscattering of Cosmic Microwave Background (CMB), Far- and Near-Infrared (FIR and NIR, respectively) radiation fields by relativistic electrons, assumed to be accelerated at the jet termination shock.
Additionally, the observed X-ray data is also simultaneously explained via the synchrotron emission from the same accelerated electrons in the background magnetic field $B$, which was considered to be a free parameter of the model.
The spectrum of the accelerated electrons, similar to protons, was considered to be a simple power-law, i.e., $Q_{\rm e}(E_{\rm e}) = Q_{\rm e, 0}(E_{\rm e}/1 \ \rm TeV)^{-{\alpha_e}}$, within $E_{\rm e, \ min} \approx 0.511 \ \rm  MeV$ ($\sim$ electron rest mass energy) and $E_{\rm e, \ max} \approx 1 \ \rm PeV$, where total electron luminosity is given by $L_{\rm e} = \int^{E_{\rm e, max}}_{E_{\rm e, min}}E_{\rm e} \ Q_{\rm e}(E_{\rm e}) \ dE_{\rm e}$. 
We adopt the energy densities of the target radiation fields as given by \cite{porter06} 
with a possible local increase factor 
of 5 (see e.g., \citealt{torres14}).
The radiation field parameters at the position of the west lobe are therefore $T_{\mathrm{CMB}} = 2.73 \ \mathrm{K}$, $\epsilon_{\mathrm{CMB}} = 0.26 \mathrm{\ eV \ cm^{-3}}$, $T_{\mathrm{FIR}} = 50 \ \mathrm{K}$, $\epsilon_{\mathrm{FIR}} = 5.3 \mathrm{\ eV \ cm^{-3}}$, $T_{\mathrm{NIR}} = 5000 \ \mathrm{K}$, and $\epsilon_{\mathrm{NIR}} = 12 \mathrm{\ eV \ cm^{-3}}$, respectively.  
Assuming the same source distance and age, we let the free parameters $\alpha_{\rm e}$, $L_{\rm e}$ and $B$ vary to fit the gamma-ray and X-ray data points via the IC and synchrotron emission, respectively.
The resulting best-fit values come out to be $\alpha_{\rm e} = 1.92 \pm 0.07$, $\mathrm{log_{10}}(L_{\rm e}/ \rm erg \ s^{-1}) = 35.81 \pm 0.16$, and $B = 36 \pm 3 \ \mu G$.
The best-fit model SED along with 1$\sigma$ error band is also shown in Fig. \ref{fig:MW_SED}.
It can be seen within the error of the observed data, the leptonic model also gives a consistent fit. 
The best-fit electron luminosity of $L_{\rm e} \approx 6.5\times 10^{35} \ \rm erg \ s^{-1}$ indicates a 0.02$\%$ CR conversion efficiency.
Additionally, the leptonic model explains the X-ray data well, while also being consistent with the radio upper limit.
This clearly implies that the leptonic scenario provides a more viable description of the multiwavelength data. 

We have also tested a hybrid leptohadronic scenario to explain the multiwavelength data.
Unfortunately, due to the relative scarcity of data in different frequencies, especially in the lower bands, the leptohadronic model parameters cannot be constrained well at this point, and the uncertainties obtained are statistically insignificant.

In reality, we expect both hadrons and leptons to contribute.
The individual viability of each of the explanations makes this a possible situation too. 

More precise radio and GeV spectra of the western lobe would help constrain the particle distributions that remain poorly determined by the present broadband modelling, allowing a stronger test of the leptonic interpretation and tighter limits on a possible hadronic contribution. For 4FGL~J1913.2+0512, continued gamma-ray monitoring could establish whether the modulation recurs. If it reappears, comparing its phase relative to the SS 433 precessional ephemeris with that of the earlier active interval would provide an additional constraint on models connecting the two sources. Molecular-line studies can provide complementary constraints on potential gas counterparts, as illustrated by the cloud detected towards 4FGL~J1913.2+0512 by \cite{oka26}. Independent distance constraints for this cloud would help assess its proposed physical association with SS 433/W50.

\section{Conclusion}

Using updated \textit{Fermi}-LAT data products, source modeling, and pulsar gating, we confirm the presence of a $\sim$162-day modulation associated with 4FGL~J1913.2+0512, with the signal concentrated mainly between 5 and 10 years after launch. No corresponding feature is found in the LAT exposure or in the window-function analysis, and nearby sources in the same ROI do not show similar behaviour, supporting the conclusion that the signal is unlikely to be produced by instrumental or mission-level systematics. Although the physical origin of this transient periodicity remains unclear, its loss over time may reflect changes in the geometry and/or efficiency of the particle transport and gamma-ray production process.

We also derived the broadband SED of the western lobe of SS~433 and modeled it with hadronic and leptonic scenarios. While both can reproduce the GeV--TeV emission, the leptonic model additionally accounts for the X-ray data with a reasonable magnetic field and energetics, making it the more compelling description of the current multiwavelength observations. A hybrid leptohadronic contribution is still possible, but present data do not constrain such a model sufficiently. Overall, our results strengthen the case for a temporally unstable, periodic gamma-ray emission from 4FGL~J1913.2+0512 and favour a predominantly leptonic origin for the western lobe emission over a purely hadronic one, although they do not exclude a hadronic contribution.

\section*{Acknowledgements}

The \textit{Fermi} LAT Collaboration acknowledges generous ongoing support from a number of agencies and institutes that have supported both the development and the operation of the LAT as well as scientific data analysis. These include the National Aeronautics and Space Administration and the Department of Energy in the United States, the Commissariat \`a l'Energie Atomique and the Centre National de la Recherche Scientifique / Institut National de Physique Nucl\'eaire et de Physique des Particules in France, the Agenzia Spaziale Italiana and the Istituto Nazionale di Fisica Nucleare in Italy, the Ministry of Education, Culture, Sports, Science and Technology (MEXT), High Energy Accelerator Research Organization (KEK) and Japan Aerospace Exploration Agency (JAXA) in Japan, and the K.~A.~Wallenberg Foundation, the Swedish Research Council and the Swedish National Space Board in Sweden.

Additional support for science analysis during the operations phase is gratefully acknowledged from the Istituto Nazionale di Astrofisica in Italy and the Centre National d'\'Etudes Spatiales in France. This work was performed in part under DOE Contract DE-AC02-76SF00515.

This work has been supported by the grant PID2024-155316NB-I00 funded by MICIU /AEI /10.13039/501100011033 / FEDER, UE and CSIC PIE 202350E189. This work was also supported by the Spanish program Unidad de Excelencia María de Maeztu CEX2020-001058-M and also supported by MCIN with funding from European Union NextGeneration EU (PRTR-C17.I1).
D.F.T. acknowledges the T. D. Lee Institute where part of this research was done, for hospitality.
A.D.S. is supported by the Juan de la Cierva JDC2023-052168-I grant, funded by MICIU/AEI/10.13039/501100011033 and by the ESF+.
Work at NRL is supported by NASA.

This publication utilizes data from Galactic ALFA HI (GALFA HI) survey data set obtained with the Arecibo L-band Feed Array (ALFA) on the Arecibo 305m telescope. The Arecibo Observatory is operated by SRI International under a cooperative agreement with the National Science Foundation (AST-1100968), and in alliance with Ana G. Méndez-Universidad Metropolitana, and the Universities Space Research Association. The GALFA HI surveys have been funded by the NSF through grants to Columbia University, the University of Wisconsin, and the University of California.

\section*{Data Availability}

All the data used in this work are publicly available or available on
request to the responsible for the corresponding observatory/facility.



\bibliographystyle{mnras}
\bibliography{ss433} 

\appendix
\section{Effect of flux reduction to periodicity significance}
\label{app:SimulatedLS}
\begin{figure}
    \centering
    \includegraphics[width=\columnwidth]{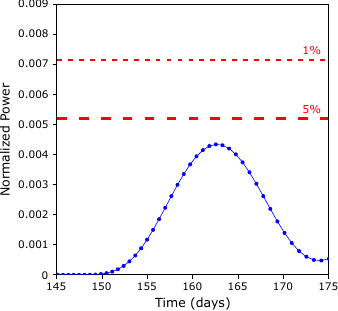}
    \caption{Lomb--Scargle periodogram of a simulated 5--10 yr data set in which the normalisation of 4FGL~J1913.2+0512 is reduced by 30 per cent (spectral shape fixed). The periodic peak persists but with reduced significance.}
    \label{fig:SimulatedLS}
\end{figure}

Table~\ref{tab:2} indicates that the $1-300$ GeV flux of 4FGL~J1913.2+0512 decreases from $(5.24 \pm 0.21) \times 10^{-12}$ \eflux in the 5--10 years interval to $(3.79 \pm 1.01) \times 10^{-12}$ \eflux in the last 6 years (a $\simeq$ 28 per cent reduction). We tested whether such a drop, by itself, could account for the disappearance of the periodic feature seen in the LS periodogram. Starting from the best-fitting model of the 5--10 years interval, we generated a simulated realisation in which the source normalisation was manually reduced by 30 per cent while the spectral shape was held fixed. We then repeated the identical LS-periodogram procedure used for the real data. The simulated LS periodogram (Fig.\ref{fig:SimulatedLS}) still shows a peak at the same period, though with a lower significance compared to the original 5--10 years interval. This suggests that the $\sim$30\% flux reduction alone is insufficient to fully eliminate the periodic signal, and cannot explain the essentially featureless LS periodogram obtained for the last 6$-$year data set (Fig.~\ref{fig:LSPeriodograms}c).

\section{Exposure}
\label{app:Exposure}
\begin{figure*}
    \centering
    \includegraphics[width=\textwidth]{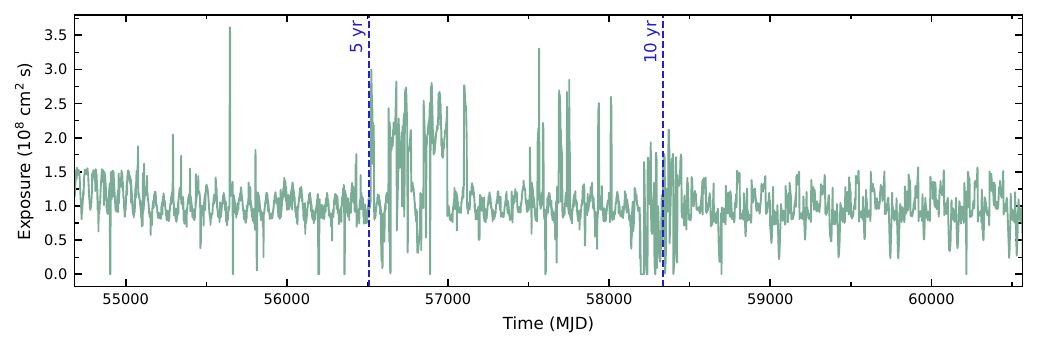}
    \includegraphics[width=\textwidth]{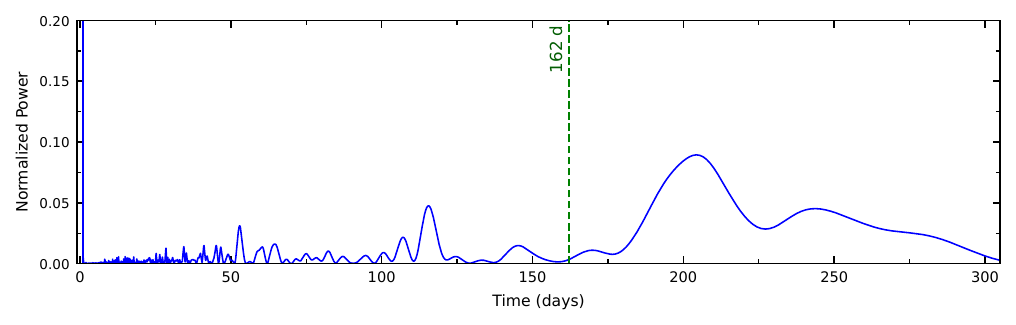}
       \caption{Top: Daily \textit{Fermi}-LAT exposure at the position of 4FGL~J1913.2+0512 from August 2008 to September 2024. Bottom: Lomb--Scargle periodogram of the daily \textit{Fermi}-LAT exposure values associated with the exposure-corrected light curve of 4FGL~J1913.2+0512 for the 5--10 years interval of the mission. The vertical dashed line marks 162 days.}
    \label{fig:Exposure}
\end{figure*}

The top panel of Fig.~\ref{fig:Exposure} shows the \textit{Fermi}-LAT daily exposure at the position of 4FGL~J1913.2+0512 from August 2008 to September 2024. 
The exposure is stable over the full interval with no significant deviations. 
The temporary high-exposure episodes around 56900 MJD are associated with periods of increased coverage during the pointed observations towards the Galactic Centre, while short-timescale fluctuations near 58200 MJD coincide with the time of the LAT solar panel anomaly. 
Integrating over the three analysis epochs yields comparable total exposures:

\begin{itemize}
    \item MJD 54683--56509 (0--5 years): $1.933 \times 10^{11}$ cm$^{2}$\,s
    \item MJD 56510--58335 (5--10 years): $2.154 \times 10^{11}$ cm$^{2}$\,s
    \item MJD 58336--60564 (10--16 years): $2.189 \times 10^{11}$ cm$^{2}$\,s
\end{itemize}

The total exposures differ by at most $\sim 10\%$ across the three time intervals, including the epoch when the source is significantly detected and the later epoch when no detection is observed, largely due to the Galactic Centre pointing strategy; however, these differences are not enough to explain the disappearance of the periodicity.
Therefore, the disappearance of the periodicity cannot be attributed to differences in exposure between intervals.

To further assess whether the $\sim$162-day modulation observed in 4FGL~J1913.2+0512 could be driven by instrumental effects, we examined the temporal behaviour of the LAT exposure itself.
For this purpose, we computed a LS periodogram of the daily exposure values associated with the exposure-corrected light curve used in the periodicity analysis. Thus, each point in the exposure time series corresponds directly to the exposure assigned to the corresponding daily flux bin.

The bottom panel of Fig.~\ref{fig:Exposure} shows this analysis for the 5--10 years interval of the mission, where the periodic signal in the gamma-ray data is strongest. 
Since the LAT exposure is governed primarily by deterministic orbital motion and survey strategy rather than stochastic noise, FAP estimates based on Gaussian white-noise assumptions are not strictly applicable. 
We therefore focus on the qualitative behaviour of the exposure periodogram rather than assigning statistical significance levels.

The exposure periodogram shows several peaks, the most prominent one, at $<$1 day and another around $\sim$53 days, consistent with known orbital and precessional periods of the spacecraft.
Additional broader features around $\sim$120 and $\sim$200 days reflect longer-term exposure modulation and/or survey strategy effects rather than coherent periodicities. 
No notable peak is observed at 162 days, suggesting that the detected signal at this period is unlikely to be driven by exposure variations.

\label{lastpage}
\end{document}